\documentclass[prd,twocolumn,nofootinbib]{revtex4}
\usepackage{bm,amsmath,amssymb,graphicx}

\begin{document}

\title{Black Hole Genesis and origin of cosmic acceleration}
\author{Nikodem J. Pop{\l}awski}
\altaffiliation{NPoplawski@newhaven.edu}
\affiliation{Department of Mathematics and Physics, University of New Haven, 300 Boston Post Road, West Haven, CT 06516, USA}


\begin{abstract}
We consider a hypothesis that the closed Universe was formed on the other side of the event horizon of a black hole existing in another universe.
That black hole appears in the Universe as a boundary white hole, and its rest frame in comoving coordinates is a frame of reference in which the cosmic microwave background is isotropic.
We consider the Lagrangian density for the gravitational field that is proportional to the curvature scalar, and use the metric-affine variational principle in which the symmetric affine connection and the metric tensor are variables.
The white hole appears in the Lagrangian through a simplest, generally covariant and linear term: the four-divergence of the four-velocity of the white hole in comoving coordinates.
We show that the variation of the action with respect to the connection generates the nonmetricity, which creates a term in the Lagrangian that is equivalent to a positive cosmological constant.
The current cosmic acceleration may therefore be a manifestation of the boundary of the closed Universe.
We also show that the equation of motion of a test particle deviates from the geodesic equation by a term that depends on the four-velocities of the particle and the white hole.
The rest frame of the white hole in comoving coordinates is the only absolutely inertial frame of reference.
This deviation might be observed on galactic scales.
\end{abstract}
\maketitle

{\it Introduction.}
A recent evidence suggests that the Universe may be closed \cite{closed}.
If the Universe was born as a closed, baby universe on the other side of the event horizon of a black hole existing in a parent universe \cite{Pat,cosmo}, then that black hole appears in the Universe as the other side of a white hole.
This white hole is the boundary of the Universe and we shall refer to it as the Parent White Hole (PWH).
In a recent, brief note, we proposed that the rest frame of PWH in comoving coordinates is a frame of reference provides the absolute inertial frame of reference (AIFR) in the Universe \cite{BHG}.
Consequently, this scenario, which we referred to as the Black Hole Genesis (BHG), could give the origin of inertia and complete Einstein's general theory of relativity by making it consistent with Mach's principle, according to which all distant matter determines inertia \cite{Mach,Lord}.
Because of the symmetry considerations, AIFR would be the frame of reference in which the average cosmic microwave background (CMB) radiation is isotropic.
Such a frame constitutes the preferred frame in the Universe, in which the cosmic time is defined \cite{Peebles}.

The Universe created by a black hole is closed, with the exception of PWH that connects the Universe to the parent universe through an Einstein--Rosen bridge.
A physical law that turns black holes into Einstein--Rosen bridges to new, baby universes must avoid the black-hole singularity.
The simplest and most natural mechanism for preventing gravitational singularities is provided by spacetime torsion within the Einstein--Cartan (EC) theory of gravity \cite{Lord,EC,Niko}.
In this theory, torsion is coupled to the intrinsic angular momentum of fermionic matter, allowing for the spin-orbit interaction that follows from the Dirac equation.
Accordingly, torsion brings the consistency between relativistic quantum mechanics and curved spacetime.
At extremely high densities, torsion manifests itself as repulsive gravity, preventing the formation of a singularity and creating a Big Bounce that starts a new, closed universe \cite{cosmo,avert} whose energy is zero \cite{energy}.
Torsion may also explain inflation \cite{cosmo}, the matter-antimatter asymmetry in the Universe \cite{anti}, or the present cosmic acceleration \cite{exp}.

{\it Lagrangian for gravitational field.}
We consider the Lagrangian density for the gravitational field that is proportional to the curvature scalar, as in the general theory of relativity \cite{Mach,LL2} or EC \cite{Lord,EC,Niko}.
AIFR appears in the Lagrangian density through a term that is linear (for simplicity) in the four-velocity vector of PWH.
To ensure locality, this term at a given point depends on the four-velocity of a body that is at rest relative to PWH in comoving coordinates and passes through this point.
We can regard this four-velocity as the four-velocity of PWH in comoving coordinates.
This four-velocity is also the four-velocity of a body that is at rest relative to CMB and passes through this point.
The simplest scalar that depends locally on this four-vector is its four-divergence.
Accordingly, we have
\begin{equation}
    \mathfrak{L}=-\frac{1}{2\kappa}(R+\theta^\rho_{\phantom{\rho};\rho})\sqrt{-g},
    \label{Lagr}
\end{equation}
where $R=R_{\mu\nu}g^{\mu\nu}$ is the curvature scalar, $\theta^\rho$ is a four-vector proportional to the aforementioned four-velocity, a semicolon denotes a covariant derivative with respect to the affine connection $\Gamma^{\,\,\rho}_{\mu\,\nu}$, $g$ is the determinant of the metric tensor $g_{\mu\nu}$ (which is reciprocal to $g^{\mu\nu}$), and $\kappa=8\pi G/c^4$.
This Lagrangian density is generally covariant.

{\it Field equations.}
We derive the field equations for the Lagrangian density (\ref{Lagr}) using the metric-affine variational principle \cite{MetricAffine}.
Since the torsion tensor (the antisymmetric part of the affine connection) is important only in the early Universe, we consider a symmetric connection: $\Gamma^{\,\,\rho}_{\mu\,\nu}=\Gamma^{\,\,\rho}_{\nu\,\mu}$.
We consider the stationary variation of the action corresponding to this Lagrangian with respect to the affine connection.
The variation of the Ricci tensor $R_{\mu\nu}=R^\rho_{\phantom{\rho}\mu\rho\nu}$ is given by $\delta R_{\mu\nu}=\delta\Gamma^{\,\,\rho}_{\mu\,\nu;\rho}-\delta\Gamma^{\,\,\rho}_{\mu\,\rho;\nu}$ \cite{Lord,Niko,LL2,nonR}.
Since $\theta^\rho_{\phantom{\rho};\rho}=\theta^\rho_{\phantom{\rho},\rho}+\Gamma^{\,\,\rho}_{\mu\,\rho}\theta^\mu$, where a comma denotes a partial derivative, the variation of the action $S=\int\mathfrak{L}\,d\Omega$ with respect to the connection gives
\begin{widetext}
\begin{eqnarray}
    & & \int(\delta\Gamma^{\,\,\rho}_{\mu\,\nu;\rho}g^{\mu\nu}-\delta\Gamma^{\,\,\rho}_{\mu\,\rho;\nu}g^{\mu\nu}+\delta\Gamma^{\,\,\rho}_{\mu\,\rho}\theta^\mu)\sqrt{-g}d\Omega \nonumber \\
    & & =\int\delta\Gamma^{\,\,\rho}_{\mu\,\nu}[-(g^{\mu\nu}\sqrt{-g})_{;\rho}+\delta^\nu_\rho(g^{\mu\sigma}\sqrt{-g})_{;\sigma}+\theta^\mu\delta^\nu_\rho\sqrt{-g}]=0,
    \label{variation}
\end{eqnarray}
\end{widetext}
where $d\Omega$ is an element of the four-volume and $\delta^\nu_\rho$ is the Kronecker tensor.
We used the integration by parts in curved space with a symmetric affine connection, with the variation vanishing on the boundary of the integration domain: $\int({\sf V}^\mu)_{;\mu}d\Omega=0$ for an arbitrary vector density ${\sf V}^\mu$ \cite{Lord,Niko,nonR}.
Since the variation of the connection $\delta\Gamma^{\,\,\rho}_{\mu\,\nu}$ is a tensor and is arbitrary, (\ref{variation}) gives the field equation:
\begin{widetext}
\begin{equation}
    -(g^{\mu\nu}\sqrt{-g})_{;\rho}+\frac{1}{2}\delta^\nu_\rho(g^{\mu\sigma}\sqrt{-g})_{;\sigma}+\frac{1}{2}\delta^\mu_\rho(g^{\nu\sigma}\sqrt{-g})_{;\sigma}+\frac{1}{2}\theta^\mu\delta^\nu_\rho\sqrt{-g}+\frac{1}{2}\theta^\nu\delta^\mu_\rho\sqrt{-g}=0.
    \label{field}
\end{equation}
\end{widetext}
Contracting the indices $\nu,\rho$ gives $(g^{\mu\sigma}\sqrt{-g})_{;\sigma}=-(5/3)\theta^\mu\sqrt{-g}$, which upon substitution into (\ref{field}) yields
\begin{equation}
    (g^{\mu\nu}\sqrt{-g})_{;\rho}=-\frac{1}{3}(\theta^\mu\delta^\nu_\rho+\theta^\nu\delta^\mu_\rho)\sqrt{-g}.
    \label{reduced}
\end{equation}
This equation shows that the affine connection in the presence of the vector $\theta^\mu$ is not metric compatible, but contains the nonmetricity tensor $g_{\mu\nu;\rho}$ that does not vanish \cite{Niko,nonR}.

A symmetric affine connection can be decomposed as
\begin{equation}
    \Gamma^{\,\,\rho}_{\mu\,\nu}=\{^{\,\,\rho}_{\mu\,\nu}\}+C^\rho_{\phantom{\rho}\mu\nu},
    \label{decompose}
\end{equation}
where $\{^{\,\,\rho}_{\mu\,\nu}\}=\frac{1}{2}g^{\rho\lambda}(g_{\nu\lambda,\mu}+g_{\mu\lambda,\nu}-g_{\mu\nu,\lambda})$ are the Christoffel symbols and $C^\rho_{\phantom{\rho}\mu\nu}$ is a tensor which is a linear combination of the nonmetricity tensors \cite{Lord,Niko,nonR}.
If a colon denotes a covariant derivative with respect to the Christoffel symbols (the metric compatible Levi-Civita connection), then we obtain, using $(\sqrt{-g})_{;\rho}=\Gamma^{\,\,\sigma}_{\sigma\,\rho}\sqrt{-g}$:
\begin{eqnarray}
    & & (g^{\mu\nu}\sqrt{-g})_{;\rho}=(g^{\mu\nu}\sqrt{-g})_{:\rho} \nonumber \\
    & & +(C^\mu_{\phantom{\mu}\sigma\rho}g^{\sigma\nu}+C^\nu_{\phantom{\mu}\sigma\rho}g^{\mu\sigma}-g^{\mu\nu}C^\sigma_{\phantom{\sigma}\sigma\rho})\sqrt{-g}.
\end{eqnarray}
Since the first term on the right-hand side of this equation vanishes, (\ref{reduced}) gives
\begin{equation}
    g^{\mu\nu}C^\sigma_{\phantom{\sigma}\sigma\rho}-C^\mu_{\phantom{\mu}\sigma\rho}g^{\sigma\nu}-C^\nu_{\phantom{\mu}\sigma\rho}g^{\mu\sigma}=\frac{1}{3}(\theta^\mu\delta^\nu_\rho+\theta^\nu\delta^\mu_\rho).
\end{equation}
The solution of this equation is
\begin{equation}
    C^\rho_{\phantom{\rho}\mu\nu}=-\frac{1}{2}\theta^\rho g_{\mu\nu}+\frac{1}{6}(\theta_\mu\delta^\rho_\nu+\theta_\nu\delta^\rho_\mu),
    \label{connection}
\end{equation}
where $\theta_\mu=g_{\mu\nu}\theta^\nu$.

Substituting (\ref{decompose}) into the curvature tensor $R^\lambda_{\phantom{\lambda}\rho\mu\nu}=\Gamma^{\,\,\lambda}_{\rho\nu,\mu}-\Gamma^{\,\,\lambda}_{\rho\mu,\nu}+\Gamma^{\,\,\sigma}_{\rho\,\nu}\Gamma^{\,\,\lambda}_{\sigma\,\mu}-\Gamma^{\,\,\sigma}_{\rho\,\mu}\Gamma^{\,\,\lambda}_{\sigma\,\nu}$ gives
\begin{eqnarray}
    & & R^\lambda_{\phantom{\lambda}\rho\mu\nu}=P^\lambda_{\phantom{\lambda}\rho\mu\nu}+C^\lambda_{\phantom{\lambda}\rho\nu:\mu}-C^\lambda_{\phantom{\lambda}\rho\mu:\nu} \nonumber \\
    & & +C^\sigma_{\phantom{\sigma}\rho\nu}C^\lambda_{\phantom{\lambda}\sigma\mu}-C^\sigma_{\phantom{\sigma}\rho\mu}C^\lambda_{\phantom{\lambda}\sigma\nu}
\end{eqnarray}
where $P^\lambda_{\phantom{\lambda}\rho\mu\nu}$ is the Riemann tensor (the curvature tensor constructed from the Levi-Civita connection) \cite{Mach,Lord,Niko,LL2,nonR}.
Consequently, the curvature scalar is given by
\begin{equation}
    R=P+C^{\mu\nu}_{\phantom{\mu\nu}\nu:\mu}-C^{\mu\nu}_{\phantom{\mu\nu}\mu:\nu}+C^{\sigma\nu}_{\phantom{\sigma\nu}\nu}C^\mu_{\phantom{\mu}\sigma\mu}-C^{\sigma\nu}_{\phantom{\sigma\nu}\mu}C^\mu_{\phantom{\mu}\sigma\nu},
\end{equation}
where $P$ is the Riemann scalar (the curvature scalar constructed from the Levi-Civita connection and the metric tensor).
Substituting (\ref{connection}) into the curvature scalar gives
\begin{equation}
    R+\theta^\rho_{\phantom{\rho};\rho}=P-\theta^\mu_{\phantom{\mu}:\mu}+\frac{1}{6}\theta^\mu\theta_\mu.
    \label{Ricci}
\end{equation}
The second term on the right-hand side of this equation is a four-divergence and does not contribute to the field equations.

{\it Cosmological constant.}
If we postulate that the vector $\theta^\mu$ is proportional to the four-velocity of PWH in comoving coordinates, $\tilde{u}^\mu$:
\begin{equation}
    \theta^\mu=\pm\sqrt{12\Lambda}\tilde{u}^\mu,
    \label{prop}
\end{equation}
then (\ref{Lagr}) and (\ref{Ricci}) give
\begin{equation}
    \mathfrak{L}=-\frac{1}{2\kappa}(P+2\Lambda)\sqrt{-g},
\end{equation}
which is the general-relativistic Lagrangian density of the gravitational field with a positive cosmological constant $\Lambda$ \cite{Lord,LL2}.
The variation of the corresponding action with respect to the metric tensor gives the Einstein field equations in the presence of the cosmological constant.
Therefore, BHG may give the origin of a positive cosmological constant, which is the simplest explanation of dark energy.
Such a constant appears in the Lagrangian as a proportionality constant that gives $\tilde{u}^\mu$ a correct unit.

The Lagrangian density (\ref{Lagr}) with the relation (\ref{prop}) is generally covariant, with local Lorentz invariance broken by a nondynamical unit timelike vector: the four-velocity of PWH in comoving coordinates.
Such a nondynamical vector field could be regarded as aether \cite{Dir}.
A different construction proposed aether as a dynamical unit timelike vector field in the metric formulation \cite{aet}.
The cosmological constant may be a manifestation of a nondynamical aether associated with PWH.

{\it Equations of motion.}
The equation of motion of a massive test particle in a Riemannian curved space is given by the geodesic equation:
\begin{equation}
    \frac{du^\rho}{ds}+\{^{\,\,\rho}_{\mu\,\nu}\}u^\mu u^\nu=0,
    \label{curved}
\end{equation}
where $u^\mu=dx^\mu/ds$ is the four-velocity of the particle and $ds=\sqrt{g_{\mu\nu}dx^\mu dx^\nu}$ is an element of the interval \cite{Mach,Lord,Niko,LL2}.
In a space with the nonmetricity, we postulate that the equation of motion is given by a similar equation:
\begin{equation}
    \frac{du^\rho}{ds}+\Gamma^{\,\,\rho}_{\mu\,\nu}u^\mu u^\nu=f^\rho,
    \label{geodesic}
\end{equation}
where $f^\rho$ is a four-vector that we will determine.
Differentiating the relation $g_{\mu\nu}u^\mu u^\nu=1$ with respect to the interval and using (\ref{geodesic}) gives
\begin{equation}
    g_{\mu\nu,\rho}u^\rho u^\mu u^\nu+g_{\mu\nu}\frac{du^\mu}{ds}u^\nu+g_{\mu\nu}u^\mu\frac{du^\nu}{ds}=0.
\end{equation}
Using (\ref{decompose}) and (\ref{geodesic}), this equation becomes
\begin{widetext}
\begin{equation}
    -2\{^{\,\,\mu}_{\rho\,\sigma}\}g_{\mu\nu}u^\nu u^\rho u^\sigma+2f^\mu g_{\mu\nu} u^\nu-2C^\mu_{\phantom{\mu}\rho\sigma}g_{\mu\nu}u^\nu u^\rho u^\sigma+g_{\mu\nu,\rho}u^\mu u^\nu u^\rho=0.
\end{equation}
\end{widetext}
The first and last terms in this equation cancel out.
Using (\ref{connection}), we obtain $f^\mu=-\theta^\mu/6$, which with (\ref{prop}) gives
\begin{equation}
    \frac{du^\rho}{ds}+\Gamma^{\,\,\rho}_{\mu\,\nu}u^\mu u^\nu=\mp\sqrt{\frac{\Lambda}{3}}\tilde{u}^\rho.
\end{equation}
Finally, substituting into this equation (\ref{decompose}) and using the relations (\ref{connection}) and (\ref{prop}) gives
\begin{equation}
    \frac{du^\rho}{ds}+\{^{\,\,\rho}_{\mu\,\nu}\}u^\mu u^\nu=\mp 2\sqrt{\frac{\Lambda}{3}}(u^\rho u_\sigma-\delta^\rho_\sigma)\tilde{u}^\sigma.
    \label{motion}
\end{equation}
Multiplying this equation by $u_\rho$ gives an identity, which is consistent with the four-acceleration being orthogonal to the four-velocity.

The sign in (\ref{motion}) is negative, which can be determined from the nonrelativistic limit of the motion of a free particle.
In a locally flat frame of reference \cite{Mach,Lord,Niko,LL2} in which PWH has a nonrelativistic comoving velocity along the $x$-axis, (\ref{motion}) gives
\begin{equation}
    \frac{du^x}{dt}\approx\mp 2c\sqrt{\frac{\Lambda}{3}}(u^x u_0 \tilde{u}^0-\tilde{u}^x)\approx\mp 2c\sqrt{\frac{\Lambda}{3}}(u^x-\tilde{u}^x).
    \label{acceler}
\end{equation}
Its solution is
\begin{equation}
    u^x=\tilde{u}^x+(u^x|_{t=0}-\tilde{u}^x)e^{\mp2\sqrt{\Lambda/3}\,ct}.
\end{equation}
As $t\to\infty$, $u^x\to\tilde{u}^x$.
The difference $u^x-\tilde{u}^x$ in (\ref{acceler}) is approximately equal to the magnitude of the peculiar velocity $v_p$ of the particle divided by $c$.
Accordingly, the acceleration of a particle arising from its motion relative to AIFR is on the order of $10^{-9}$ m/s$^2 (v_p/c)$.
The numerical factor in this order of magnitude is similar to the acceleration scale that may exist in galaxies, and the resulting acceleration is peculiar-velocity-emergent and thus different for different galaxies \cite{acc}.
Similarly, one can show that $u^y\to0$ and $u^z\to0$.
Consequently, a particle with an initial peculiar motion (with respect to AIFR) asymptotically comes to rest (in comoving coordinates) relative to PWH: it tends to an inertial state.
If the sign in (\ref{motion}) were positive, then the peculiar motion of the particle would unreasonably grow in time.

Therefore, the equation of motion of a test particle in a curved space is
\begin{equation}
    \frac{du^\rho}{ds}+\{^{\,\,\rho}_{\mu\,\nu}\}u^\mu u^\nu=-2\sqrt{\frac{\Lambda}{3}}(u^\rho u_\sigma-\delta^\rho_\sigma)\tilde{u}^\sigma.
    \label{deviation}
\end{equation}
Its nonrelativistic limit in a locally flat (Galilean) frame of reference can be written in a vector form as
\begin{equation}
    \frac{d{\bf v}}{dt}\approx {\bf a}_g- 2c\sqrt{\frac{\Lambda}{3}}({\bf v}-\tilde{{\bf v}}),
\end{equation}
in accordance with (\ref{acceler}), where ${\bf a}_g$ is the gravitational acceleration.
The acceleration of a particle in this limit has a term proportional to its peculiar velocity ${\bf v}-\tilde{{\bf v}}$.
The velocity $\tilde{{\bf v}}$ is the velocity of a body that is at rest relative to PWH in comoving coordinates (and at rest relative to CMB) and passes through the location of the particle.

{\it Final remarks.}
The equation of motion (\ref{deviation}) deviates from the geodesic equation at large scales, on the order of $1/\sqrt{\Lambda}$.
Also, it reduces to the free motion in curved space (\ref{curved}) only if $u^\mu=\tilde{u}^\mu$, that is, in AIFR.
Consequently, AIFR is the only inertial frame of reference.
Other frames, moving with constant velocities with respect to AIFR \cite{LL1}, are approximately inertial: the effects of the noninertiality are significant only at large scales.
Deviations from the geodesic motion and the peculiar-velocity-emergent acceleration scale might be observed in galactic motion.

We note that a simple extension of the general theory of relativity, without changing the linearity of the Lagrangian density for the gravitational field with respect to $R$ or introducing new hypothetical fields, may explain the dynamics of the early and late Universe.
Torsion, generated by the intrinsic angular momentum of elementary particles, prevents the formation of gravitational singularities and provides a physical mechanism for BHG.
Nonmetricity, generated by the boundary of the Universe (PWH), may explain the cosmological constant that causes the present cosmic acceleration.

This work was funded by the University Research Scholar program at the University of New Haven.

\end{document}